\documentclass[a4paper,11pt]{article}
\usepackage{pos}

\usepackage[sort&compress]{natbib}
\setcitestyle{numbers}
\usepackage{physics}
\usepackage{cleveref}
\usepackage{todonotes}
\usepackage{amsmath}
\usepackage{bbold}
\usepackage{mathtools}

\graphicspath{
	{.} 
	{./figs/}
}

\title{Lattice investigation of the general Two Higgs Doublet Model with $SU(2)$ gauge fields}

\renewcommand{\printHeadAuthors}{G. Catumba}
\author*[a]{Guilherme Catumba}
\author[h]{Atsuki Hiraguchi}
\author[c]{George W.-S Hou}
\author[d]{Karl Jansen}
\author[c,e]{Ying-Jer Kao}
\author[b,f,g]{C.-J. David Lin}
\author[a]{Alberto Ramos}
\author[c]{Mugdha Sarkar}

\affiliation[a]{Instituto de Física Corpuscular (IFIC) CSIC - Universitat de Valencia. \\ 46071, Valencia, Spain}
\affiliation[b]{Institute of Physics, National Yang Ming Chiao Tung University, 1001 Ta-Hsueh Road, Hsinchu 30010, Taiwan}
\affiliation[c]{Department of Physics, national Taiwan University, Taipei 10617, Taiwan}
\affiliation[d]{Deutsches Elektronen-Synchrotron DESY, Platanenallee 6, 15738 Zeuthen, Germany} \affiliation[e]{Center for Theoretical Physics and Center for Quantum science and technology, National Taiwan University, Taipei, 10607, Taiwan}
\affiliation[f]{Center for High Energy Physics, Chung-Yuan Christian University, Chung-Li 32023, Taiwan}
\affiliation[g]{Centre for Theoretical and Computational Physics, National Yang Ming Chiao Tung University, 1001 Ta-Hsueh Road, Hsinchu 30010, Taiwan}
\affiliation[h]{CCSE, Japan Atomic Energy Agency, 178-4-4, Wakashiba, Kashiwa, Chiba 277-0871, Japan}

\emailAdd{gtelo@ific.uv.es}

\abstract{

  We study the most general Two Higgs Doublet Model with
  $SU(2)$ gauge fields on the lattice.
  The phase space is probed through the computation of gauge-invariant global
  observables serving as proxies for order parameters.
  In each phase, the spectrum of the theory is analysed for different combinations
  of bare couplings and different symmetry breaking patterns.
  The scale setting and determination of the running gauge coupling are performed through
  the Wilson flow computation of the action density.
}

\FullConference{%
The 40th International Symposium on Lattice Field Theory,\\
31st July-4th August, 2023,\\
Fermilab, Batavia, Illinois, USA
}

\graphicspath{
  {.} 
}

\begin{document}
\maketitle

\section{Introduction} \label{sec:Introduction}

The Standard Model (SM) is a well-established but incomplete theory describing interactions among elementary particles. Extensions for physics beyond the Standard Model (BSM) attempt to tackle some of the unaddressed aspects of this theory.
With one doublet established, the simplest extension to the theory is the addition of an extra $SU(2)$ scalar doublet -- Two Higgs Doublet Model (2HDM) \cite{PhysRevD.8.1226}.
One of the motivations for this  modification comes from the presence of multiple families of quarks and leptons in the SM.
The addition of extra particles in the spectrum may come along with new features such as new sources for CP-violation, dark-matter and axion-model applications, changes in the electroweak baryogenesis, among others -- see \cite{branco_theory_2012}.
Furthermore, from supersymmetry-inspired 2HDMs come lower bounds to the masses of the extra scalar particles \cite{Athron_2017}.

While the additional features are of interest, it is fundamental that all experimentally sustained properties of the SM are kept.
Such a requirement can result in constraints on the 2HDM, such as bounds on the masses of non-SM scalar particles.
This is what we will explore in this research programme.
The appearance of flavour changing neutral currents (FCNC) and large violations of the custodial symmetry requires either a restriction to certain regions of the parameter space, or additional symmetries \cite{BhupalDev:2017txh} to be imposed on the 2HDM scalar potential (reducing the number of free parameters), \textit{e.g.}, a discrete $\mathbb Z_{2}$ symmetry for one of the doublets eliminates tree-level FCNC \cite{PhysRevD.15.1958,PhysRevD.15.1966}.
In \cite{HOU1992179}, however, it was claimed that $\mathbb Z_{2}$-symmetry may not be required to control the FCNC if certain mass-mixing hierarchies are found in the fermion flavour sector, with an additional \cite{Hou_2018} alignment of the CP-even Higgs mass eigenstates. This in turn may correspond to $\order{1}$ quartic couplings, that are also required for first order electroweak phase transition.

The possible need of large couplings, combined with the triviality  of the scalar theory, requires a careful examination of the cutoff dependence of the couplings and the spectrum.
This calls for a fully non-perturbative study of the theory.
In this work we present a preliminary investigation of the 2HDM model on the lattice, namely, we study different patterns of the symmetry breaking through the analysis of the spectrum and the running gauge coupling.

Considering only real couplings, the most general potential for the 2HDM is
\begin{align}
  \label{eq:cont potential}
  V_{\text{2HDM}} &= \mu_{11}^{2}\phi_{1}^{\dagger}\phi_{1} + \mu_{22}^{2}\phi_{2}^{\dagger}\phi_{2} + \mu_{12}^{2}\Re (\phi_{1}^{\dagger}\phi_{2}) + \lambda_1 (\phi_{1}^{\dagger}\phi_{1})^2 + \lambda_2 (\phi_{2}^{\dagger}\phi_{2})\nonumber \\
  &+ \lambda_3(\phi_{1}^{\dagger}\phi_{1})(\phi_{2}^{\dagger}\phi_{2})
  + \lambda_4(\phi_{1}^{\dagger}\phi_{2})(\phi_{2}^{\dagger}\phi_{1}) + \lambda_{5} \Re(\phi_{1}^{\dagger}\phi_{2})^{2}\nonumber \\
  &  +\Re(\phi_{1}^{\dagger}\phi_{2})\left[ \lambda_{6} (\phi_{1}^{\dagger}\phi_{1}) + \lambda_{7} (\phi_{2}^{\dagger}\phi_{2})\right]
\end{align}
where $\phi_{1,2}$ are complex scalar doublets.

On the lattice, we use the quaternion representation of the scalar fields $\Phi_{n}(x) = \frac{1}{\sqrt{2}}\sum_{a=1}^{4}\theta_{a}\varphi_{n}^{a}(x)$ where $n=1,2$ labels the scalar field, $\varphi^{a}\in\mathbb R$, while $\theta^{4}=\mathbb{1}_{2\times2},~\theta^{j}=i\sigma^{j},~j=1,2,3$ with $\sigma^{j}$ being the Pauli matrices.
In the following, we take the superscript $j$ to take values $j=1,2,3$ while the superscript $a$ ranges from 1 to 4.
Notice that there is a one-to-one correspondence between the doublet and quaternion formulation, at the level of the real components.

The full lattice $SU(2)$ gauge-invariant action in this work contains the standard Wilson plaquette action, $S_{\text{YM}}$, the gauge-Higgs interaction and the scalar potential.
\begin{align}
  \label{eq:lattice action}
  S_{\text{2HDM}} &= S_{\text{YM}}+ \sum_{x}\sum_{n=1}^{2} \bigg\{ \sum_{\mu} -2 \kappa_{n}\Tr \left( \hat\Phi_{n}^{\dagger}U_{\mu}\hat\Phi_{n}(x+\hat\mu) \right)+  \Tr \left( \hat\Phi_{n}^{\dagger}\hat\Phi_{n} \right) + \eta_{n}\left[ \Tr \left( \hat\Phi_{n}^{\dagger}\hat\Phi_{n} \right) - 1 \right]^{2}\bigg\}  \nonumber \\
  &  + 2\mu^{2}\Tr \left( \hat\Phi_{1}^{\dagger}\hat\Phi_{2} \right) + \eta_{3}\Tr \left( \hat\Phi_{1}^{\dagger}\hat\Phi_{1} \right)\Tr \left( \hat\Phi_{2}^{\dagger}\hat\Phi_{2} \right) + \eta_{4}\left[\Tr \left( \hat\Phi_{1}^{\dagger}\hat\Phi_{2} \right)^{2}+\Tr \left( \hat\Phi_{1}^{\dagger}\hat\Phi_{2} i\sigma_{3}\right)^{2} \right] \nonumber\\
  &+ \eta_{5}\left[\Tr \left( \hat\Phi_{1}^{\dagger}\hat\Phi_{2} \right)^{2}-\Tr \left( \hat\Phi_{1}^{\dagger}\hat\Phi_{2} i\sigma_{3}\right)^{2} \right] + 2  \Tr \left( \hat\Phi_{1}^{\dagger}\hat\Phi_{2} \right) \bigg[\eta_{6}\Tr \left( \hat\Phi_{1}^{\dagger}\hat\Phi_{1} \right) + \eta_{7}\Tr \left( \hat\Phi_{2}^{\dagger}\hat\Phi_{2} \right) \bigg],
\end{align}
where $U_{\mu}$ is the $SU(2)$ gauge link.
The lattice fields and couplings are related to the continuum ones by $\Phi_{n} = a\phi_{n}/\sqrt{\kappa_{n}}$, $a^{2}\mu_{nn}^{2}=(1-\eta_{n} - 8\kappa_{n})/\kappa_{n}$, $a^{2}\mu_{12}^{2}=\mu^{2}$, $\lambda_{n}=\eta_{n}/\kappa_{n}^{2},~n=1,2$, $\lambda_{n}=\eta_{n}/(\kappa_{1}\kappa_{2}),~n=3,4,5$ and, $\lambda_{6}=\eta_{6}/(\kappa_{1}^{3/2}\kappa_{2}^{2})$, $\lambda_{7}=\eta_{7}/(\kappa_{1}^{2}\kappa_{2}^{3/2})$.

\section{Phase structure, spectrum \& Gradient Flow} \label{sec:phase structure}

Contrary to the single Higgs doublet model, the addition of a second doublet guarantees the existence of two different phases in the theory, separated by a phase transition \cite{wurtz_effect_2009,lewis_spontaneous_2010}.
The system comprehends a symmetric phase with confinement-like properties and massless gauge bosons; and a broken or Higgs phase where the gauge bosons acquire mass through the Fröhlich-Morchio-Strocchi (FMS) mechanism \cite{FROHLICH1980249}.

The 2HDM may have various global symmetries depending on the choice of couplings.
An important case is the discrete $\mathbb Z_{2}$-symmetry, under which the fields change sign while leaving the action invariant.
Under the assumption of a $\mathbb Z_{2}\times \mathbb Z_{2}$-symmetric scalar potential ($\mu^{2}=\eta_{6}=\eta_{7}=0$), and following ref. \cite{deshpande_pattern_1978}, it is useful to define four `mutually exclusive' sectors of the Higgs potential by the vacuum expectation value (VEV), $v_{i}$, for each scalar field: $\textrm{(A)}: v_{1}=v_{2}=0,~\textrm{(B)}: v_{1}=0, v_{2}\neq0,~\textrm{(C)}: v_{1}\neq0, v_{2}=0,~\textrm{(D)}: v_{1}\neq0, v_{2}\neq0$.
While this is not a gauge-invariant statement it does help classifying the regions of parameter space by studying which global symmetries remain in each sector.

In order to analyse the phase structure of this theory we compute observables that serve as proxies for order parameters.
There are different observables that can be used to probe the phase transition, such as the average plaquette, $P = \sum_{x,\mu,\nu}\Tr U_{\mu\nu}$, the squared Higgs-field length, $\expval{\rho_{n}^{2}}=\expval{\sum_{x}\det(\Phi_{n}(x))}/V$, or the gauge-invariant link
\begin{equation}
  L_{\alpha_{n}} =\frac{1}{8V}\sum_{x,\mu}\Tr{\alpha_{n}^{\dagger}(x)U_{\mu}(x)\alpha_{n}(x+\hat\mu)},
\end{equation}
where $\alpha_{n}$ is the `angular' part of the quaternion Higgs field, $\Phi_{n}=\rho_{n}\alpha_{n},~\rho_{n}\in\mathbb R,~\alpha_{n}\in SU(2)$.
We note that $L_{\alpha_{n}}$ is a bounded quantity for all values of the couplings.


To study the spectrum of the theory we consider gauge-invariant composite operators (projected to zero momentum)
\begin{align}
    &S_{ij}^{a}(x^{4}) = \sum_{\vec x} \Tr\left[\Phi_i^\dagger(x)\Phi_j(x)\theta^{a}\right],\label{eq:Sij}\\
    &W_{ij,k}^a(x^{4}) = \sum_{\vec x} \Tr\left[\Phi_i^\dagger(x)U_k(x)\Phi_j(x+k)\theta^{a}\right].\label{eq:Wij}
\end{align}
that can be classified according to the representations of the global symmetry group \cite{maas_gauge_2016} as well as of the lattice $H(4)$ symmetry group.
Additionally, for the vector interpolators, the mass does not depend on the superscript $j$, and we average over this index.
In building the interpolators we perform gradient (Wilson) flow  on the gauge links  \cite{Luscher_2010_flow}, and carry out smearing on the scalar fields \cite{PhysRevD.80.054506}.
The spectrum is then obtained by computing the effective masses from the relevant correlation functions, $C(t)=\expval{O(t)O(0)} - \expval{O(t)}\expval{O(0)}$, where $O$ is one of the interpolators defined in \cref{eq:Sij,eq:Wij}.

At leading order in the lattice spacing, the quantum numbers $J^{P}$ for these operators are \cite{Wurtz:2015zsz}
\begin{align}
  S_{ii}^{4},~W_{ii}^{4} \rightarrow 0^{+}, && W_{ij}^{4} \rightarrow 0^{+}~(i\neq j), && W_{ii}^{j} \rightarrow 1^{-}.
\end{align}
The connection between the interpolators and the spectrum of the theory is not straightforward.
Firstly, the interpretation of the extracted masses from these operators depends on the particular phase of the system.
Furthermore, the squared-mass matrices in the broken phases are not, in general, diagonal when written with the original fields.
For this case a variational method is required to isolate the mass-eigenstates.
In \cite{PhysRevD.74.015018} the mass matrices were computed for a general basis.
In the Inert limit $\mu^{2}=\eta_{6}=\eta_{7}=0$, and real couplings ($\mathbb Z_{2}\times \mathbb Z_{2}$-symmetric theory)  there is no mixing if one chooses to work with the Higgs basis, where only one of the scalar fields acquires a vacuum expectation value.
For simplicity, and since the Inert model condition is stable under renormalization, in the following we consider $\mu^{2}=\eta_{6}=\eta_{7}=0$.

Under the assumption that the mass matrix is diagonal we can establish the connection between the mass-eigenstates and the interpolators for the symmetric and broken phases.
In the symmetric phase the spectrum of the theory accomodates two distinct scalar particles --   the correlators of $S_{ii}^4$ and $W_{ii}^4$ both yield ground-state energies $2m_{i}$.
The mixed interpolators  $S_{12}^a$ and $W_{12}^a$ both couple to the same scalar state whose lowest energy is $\sim (m_{1}+m_{2})$.
The massless gauge vector bosons appear in the spectrum of $W_{ii}^{j},~i=1,2$.

In the broken phase corresponding to sector (B), $v_{1}=0, v_{2}\neq0$, realised in the regime  $\kappa_{1}<\kappa_{1}^{\textrm{crit}}$ and $\kappa_{2}>\kappa_{2}^{\textrm{crit}}$,  the gauge bosons become massive and there may be up to five distinct scalar particles\footnote{
Heuristically, from the usual expansion around the non-vanishing minimum of the scalar potential, this is equivalent to only $\Phi_{2}$ acquiring a non-zero vacuum expectation value, \textit{i.e.}, sector (B). Considering the expansion around this expectation value, $v$, the mass eigenstates can be written \cite{PhysRevD.74.015018} in the Inert case as
\begin{align*}
    &m_{h}^{2} = \eta_{2}v^{2}, && m_{H^{\pm}}^{2} = \mu_{11}^{2}+\eta_{3}v^{2}/2, \\
    &m_{A}^{2} = m_{H^{\pm}}^{2} + (\eta_{4} - \eta_{5})v^{2}/2, && m_{H}^{2} = m_{A}^{2} + \eta_{5}v^{2} = m_{H^{\pm}}^{2} + (\eta_{4} + \eta_{5})v^{2}/2
\end{align*}
In the custodial symmetric limit $\eta_{4}=\eta_{5}$ there is a scalar degenerate triplet.
}.
It can be shown  \cite{maas_gauge_2016} that for this choice both $S_{22}^4$ and $W_{22}^4$ yield the mass of the elementary Higgs, $m_{h}$, while the vector operator $W_{22}^{j}$ sources the degenerate triplet associated with the $W$-bosons.
The mixed scalar operators $S_{12}^a$ and $W_{12}^a$ both couple to the four components of the second scalar doublet representing the additional BSM scalar particles with masses $m_{H}\equiv m^{(4)}$ and $m^{(j)},~j=1,2,3$.
Finally, the vector operator $W_{11}^{j}$ couples to a heavy excited stated.

As an exploratory work, we are interested in learning how to control the bare couplings of this theory such that we can later impose physical conditions.
In addition to tuning the Higgs-to-$W$ mass ratio to its physical value, we impose the condition such that the value of the running gauge coupling at the W-boson mass scale takes its physical value.
To this end, we work with the renormalization scheme defined through the Gradient Flow action density \cite{Luscher_2010_flow}, $\expval{E(x,t)} = -\frac{1}{4}\expval{G_{\mu\nu}^{a}(x,t)G_{\mu\nu}^{a}(x,t)}$, where $G_{\mu\nu}(x,t)$ is the flowed gauge field strength at flow time $t$ (we consider the Clover discretization).
Using the relation between the renormalized gauge coupling at the scale $\mu = 1/\sqrt{8t}$ and the flowed action density \cite{luscher_perturbative_2011} we define the Gradient Flow renormalized gauge coupling by
\begin{align}
    g_{GF}^{2}(\mu)\equiv \frac{128\pi^{2}}{9} t^{2}\expval{E(t)}\eval{}_{t=1/8\mu^{2}}, && g_{GF}^{2}(\mu = m_{W}) = 0.5.
\end{align}

For the lattice simulations, the Monte Carlo configurations were generated using the Hybrid Monte Carlo algorithm.
The code implementation for NVIDIA GPUs can be found in \cite{ramos_latgpu}.
All results shown in this article are obtained from $24^{4}$ lattices (volume effects were not observed when considering a $32^{4}$ lattice).

\section{Results} \label{sec:Results}

Due to the large number of quartic couplings for the scalar fields in \cref{eq:cont potential,eq:lattice action}, various possible global symmetries can emerge depending on the constraints imposed on the scalar potential. \cite{deshpande_pattern_1978,PILAFTSIS2012465,maas_gauge_2016}.
Imposing no constraint on these couplings leads to the general scenario of an $SU(2)$ global symmetry.
As a preliminary study, we take $\eta_{4}=\eta_{5}$ which makes the scalar potential $O(4)\sim SU(2)\times SU(2)$-invariant, implying an $SU(2)$ custodial symmetry \cite{PhysRevD.74.015018} and the presence of three degenerate scalar particles.
While we intend to explore the possibility of having some $\order{1}$ quartic couplings, for the first step in our investigation we also take all quartic couplings to be small.

In order to study the symmetry breaking pattern we fix all bare couplings such that $\kappa_{1}$ is below its critical value, and induce the phase transition by increasing $\kappa_{2}$.
This corresponds to the transition between sector (A) and (B) introduced in \cref{sec:phase structure}.
Additionally, for small quartic couplings the tree-level relations for the masses give us hints on how to tune the bare couplings such that we can control the mass ratios in the theory.

\begin{figure}[htb!]
  \centering
    \begin{minipage}[t]{.5\textwidth}
		\centering
		\scalebox{0.5}{\includegraphics[scale=0.7]{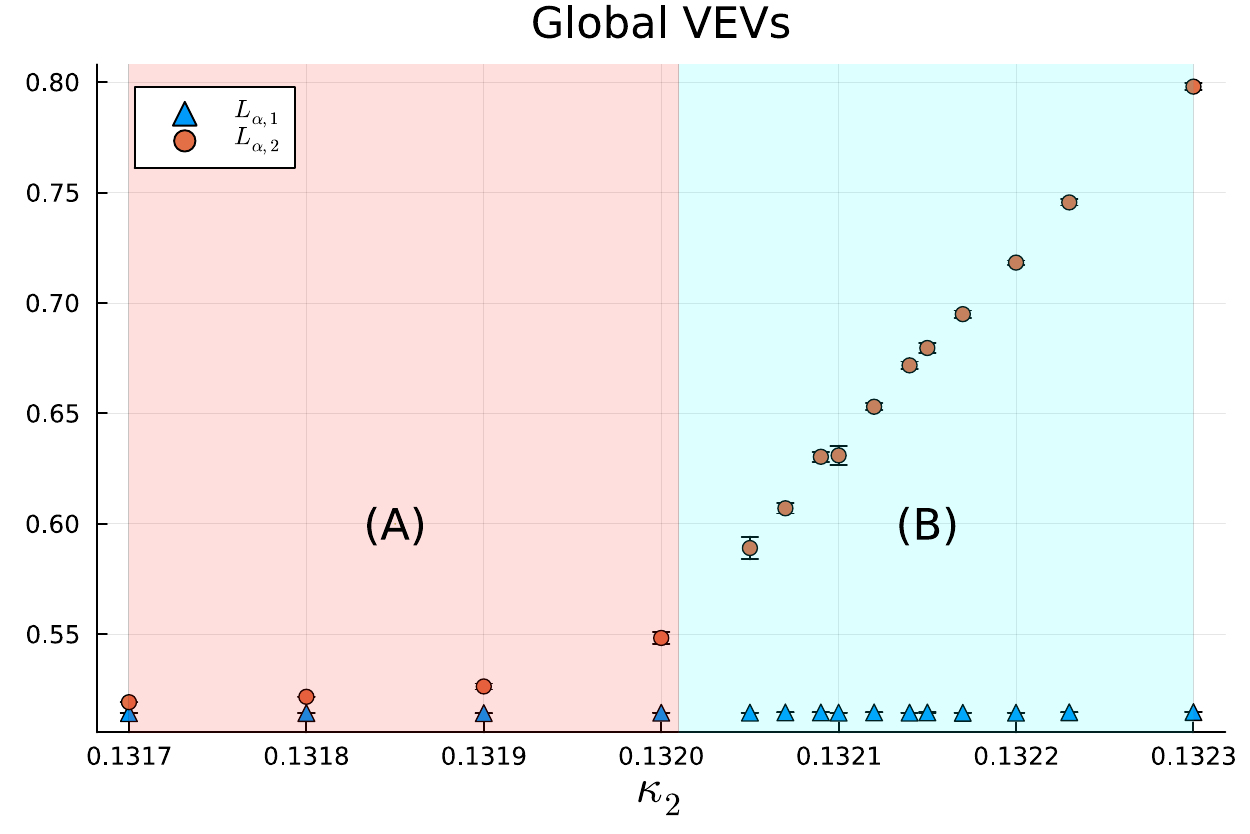}}
	\end{minipage}%
    \begin{minipage}[t]{.5\textwidth}
		\centering
		\scalebox{0.5}{\includegraphics[scale=0.7]{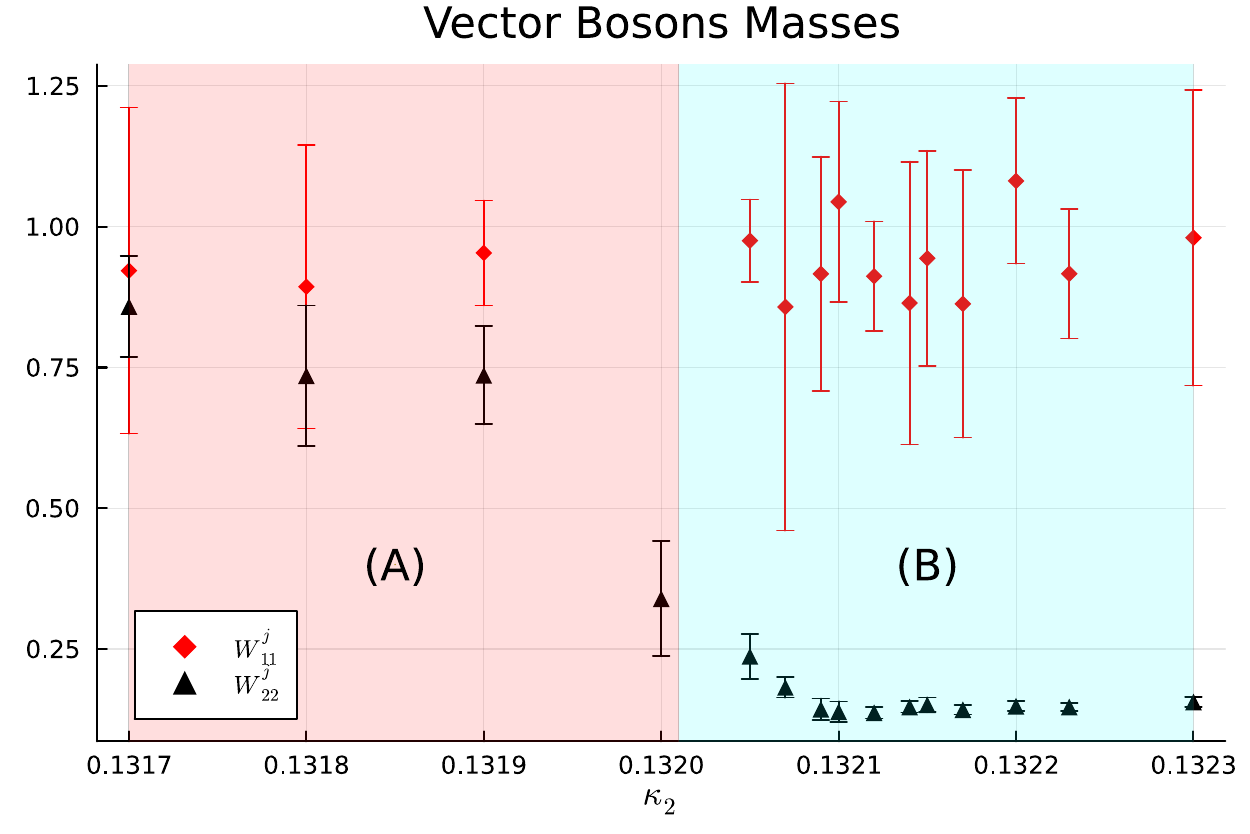}}
	\end{minipage}%
    \\
    \begin{minipage}[t]{.5\textwidth}
		\centering
		\scalebox{0.5}{\includegraphics[scale=0.7]{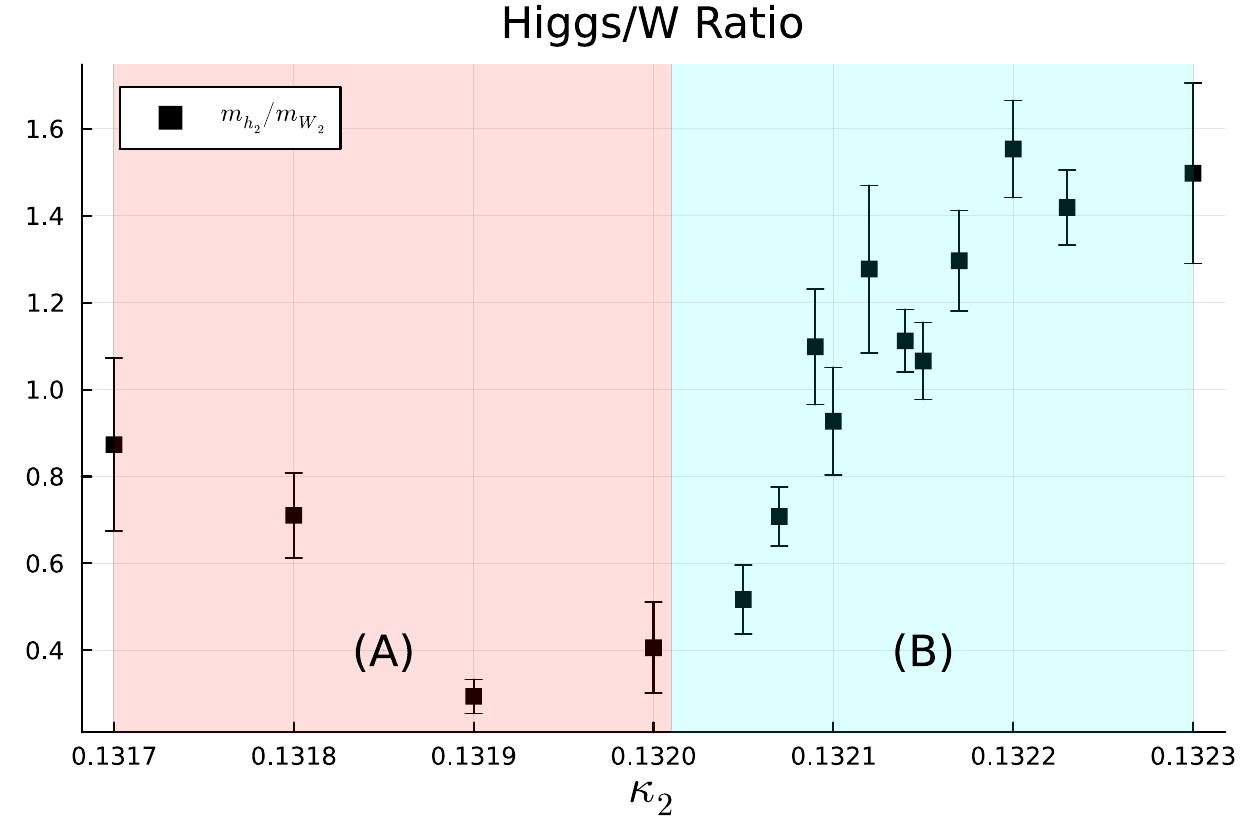}}
	\end{minipage}%
    \begin{minipage}[t]{.5\textwidth}
		\centering
		\scalebox{0.5}{\includegraphics[scale=0.7]{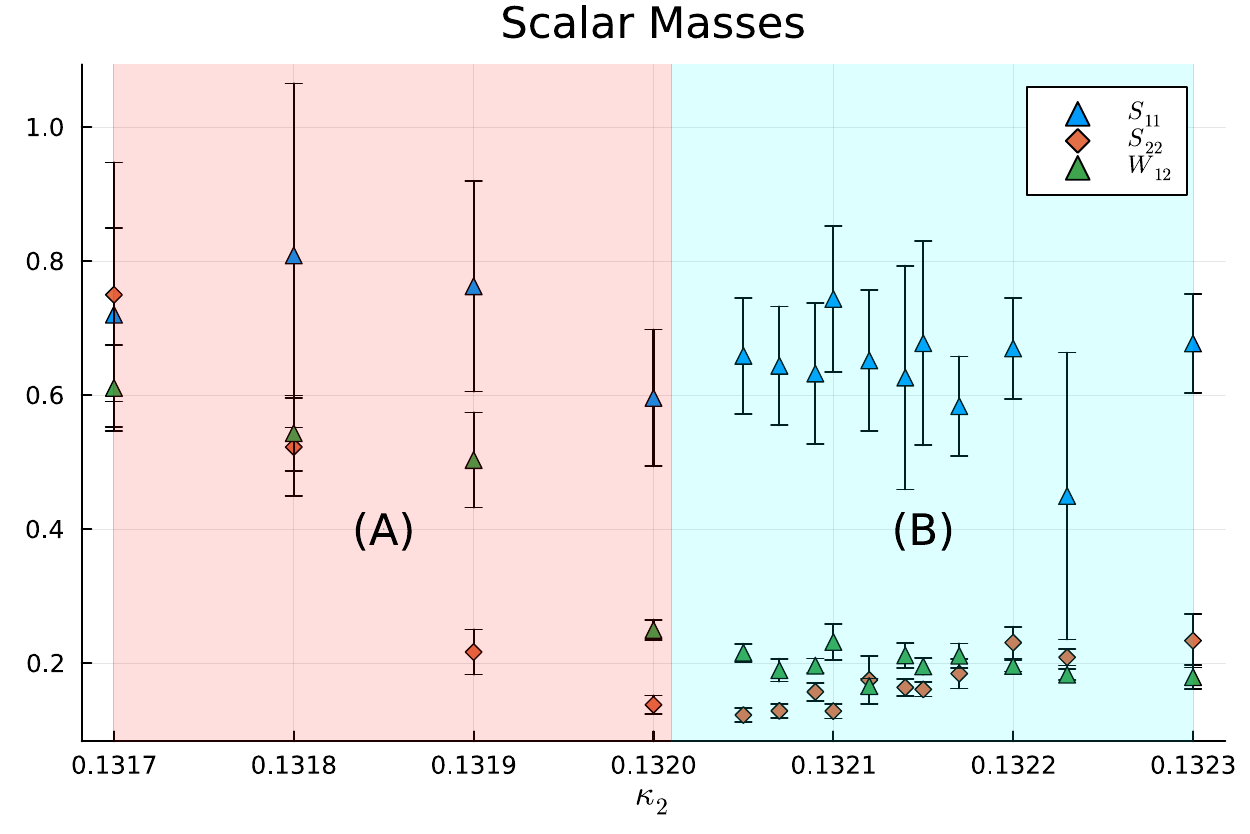}}
	\end{minipage}%
  \caption{ Transition (A) to (B) with global symmetry breaking $O(4)\times \mathbb{Z}_{2}\times \mathbb{Z}_{2} \rightarrow O(3)\times\mathbb{Z}_{2}$.
    The masses in lattice units are shown in the right-hand plots for the vector boson states (top) and the scalar states (bottom).
    Couplings: $\beta = 6.0$; $\kappa_{1}=0.1308$; $\eta_1=0.003$; $\eta_2=0.004$; $\mu=0.0$; $\eta_3=0.0001$; $\eta_4=0.00005$; $\eta_5=0.00005$; $\eta_6=0.0$; $\eta_7=0.0$. \\ }
	\label{fig:bet = 6.0; k1=0.1308; et1=0.003; et2=0.004; mu=0.0; xi1=0.0001; xi2=0.00005; xi3=0.00005;; xi4=0.0; xi5=0.0}
\end{figure}

In \cref{fig:bet = 6.0; k1=0.1308; et1=0.003; et2=0.004; mu=0.0; xi1=0.0001; xi2=0.00005; xi3=0.00005;; xi4=0.0; xi5=0.0} we show how various quantities change across the transition from sectors (A) to (B) corresponding to the global symmetry breaking pattern $O(4)\times \mathbb{Z}_{2}\times \mathbb{Z}_{2} \rightarrow O(3)\times\mathbb{Z}_{2}$.
In the top-left plot the gauge-invariant link $\expval{L_{\alpha_{2}}}$ signals the phase transition, rapidly increasing for $\kappa_{2}\gtrsim0.132$ while $\expval{L_{\alpha_{1}}}$ remains small and constant.
The lowest state energy (in lattice units) associated with the vector operators $W_{ii}^{j},~i=1,2$ are shown in the top-right plot.
While $W_{22}^{j}$ couples to the massive $W$-bosons (in the Higgs phase), the correlation functions for $W_{11}^{j}$ show a fast decay, representing a heavy scattering state.
On the bottom-right plot we show the lowest energies for the scalar operators $S_{11}^{4},~S_{22}^{4}$, and $W_{12}^{a}$, which correspond to $2m_{H}$, $m_{h}$, and $m^{(j)}$, respectively\footnote{The masses $m^{(j)},~j=1,2,3$ are degenerate as expected from the remaining $O(3)$ symmetry.}.
Finally, the ratio between the lowest energies from $S_{22}$ and $W_{22}$ is shown in the bottom-left corner.
This quantity corresponds to the Higgs-to-$W$ mass ratio in the Higgs phase, and it can be seen that, within the current precision, its value is close to the physical ratio.

\begin{figure}[htb!]
  \centering
      \begin{minipage}[t]{.5\textwidth}
		\centering
		\scalebox{0.5}{\includegraphics[scale=0.7]{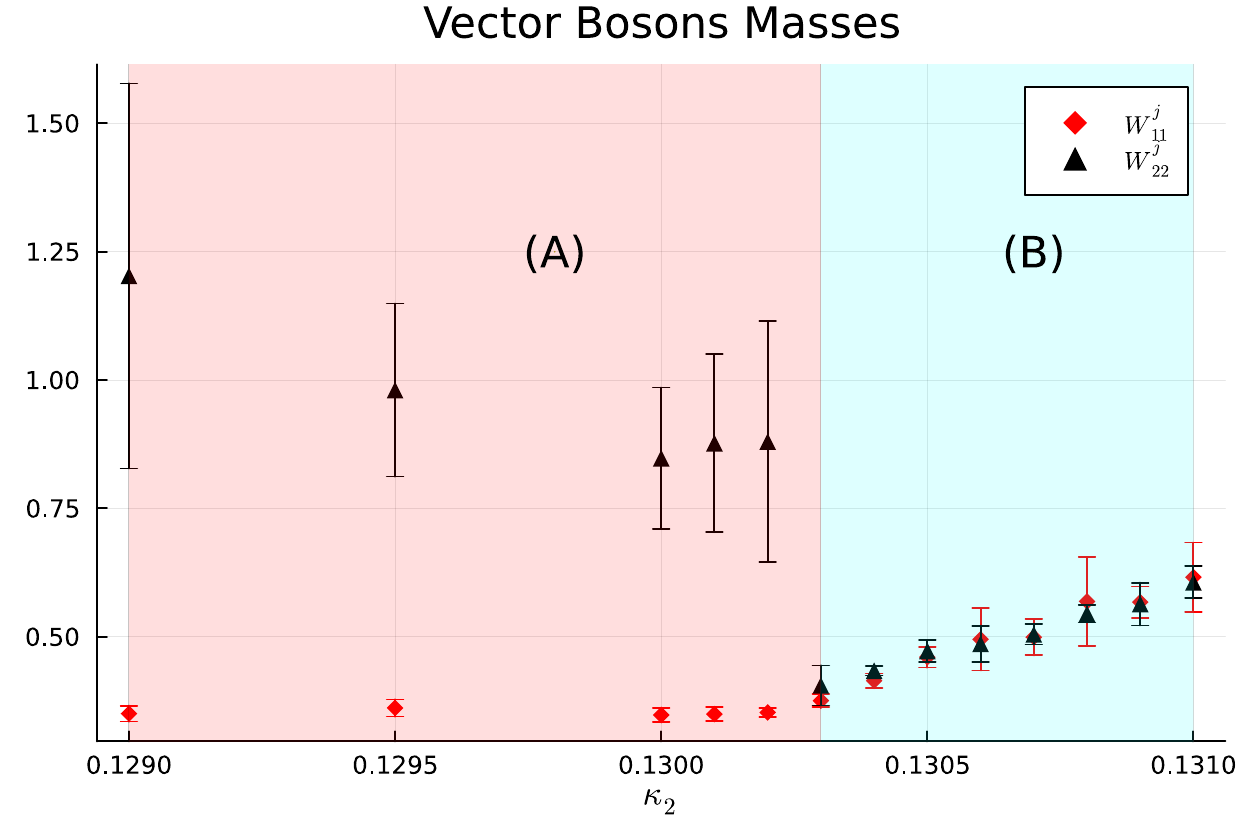}}
	\end{minipage}%
    \begin{minipage}[t]{.5\textwidth}
		\centering
		\scalebox{0.5}{\includegraphics[scale=0.7]{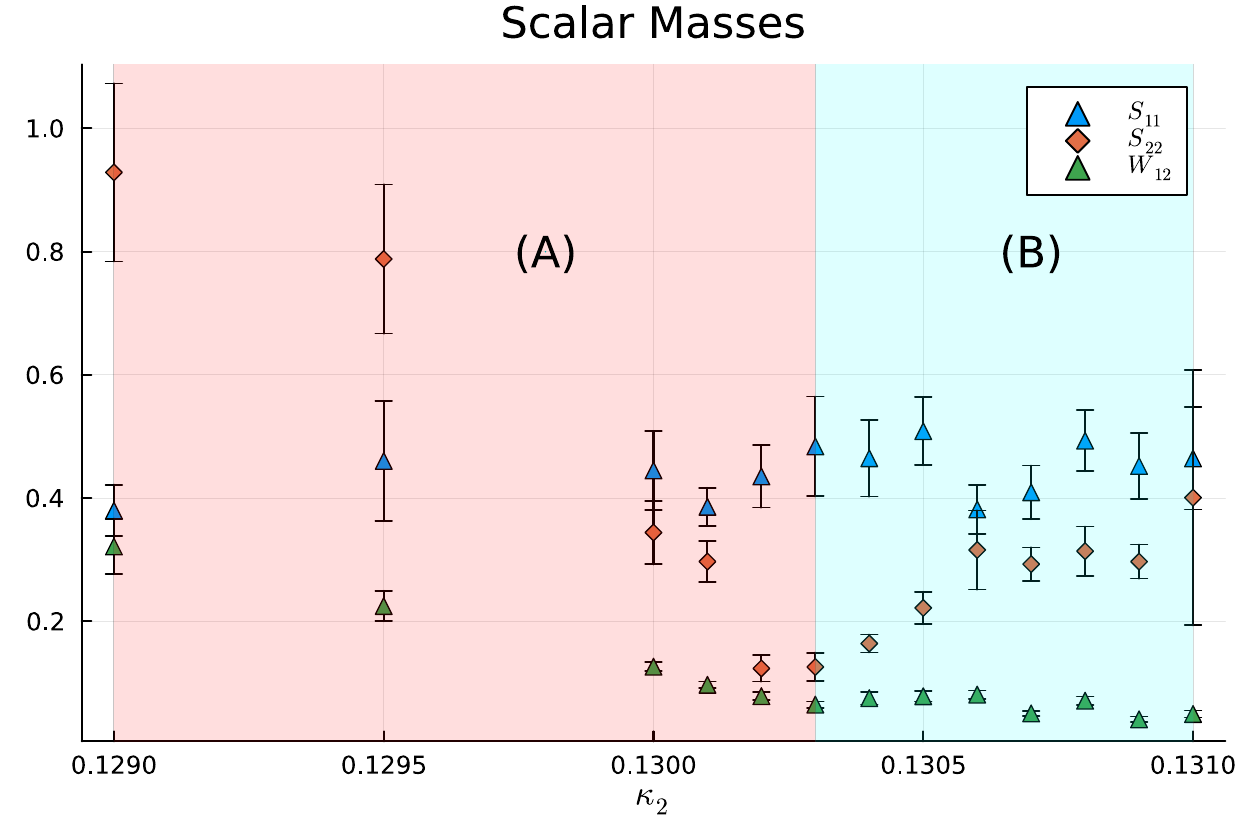}}
	\end{minipage}%
  \caption{Transition (C) to (D) with global breaking pattern $O(3)\times \mathbb{Z}_{2} \rightarrow O(2)$.
    The masses in lattice units are shown for the vector boson states (left) and the scalar states (right).
    Couplings: $\beta = 6.0$; $\kappa_{1}=0.133$; $\eta_1=0.003$; $\eta_2=0.001$; $\mu=0$; $\xi_1=0.0001$; $\xi_2=0.00005$; $\xi_3=0.00005$; $\xi_4=0$; $\xi_{5}=0$.}
	\label{fig:bet = 5.5; k1=0.133; et1=0.003; et2=0.001; mu=0.0; xi1=0.0001; xi2=0.00005; xi3=0.00005;xi4=0.0; xi5=0.0}
\end{figure}

A complementary test to the symmetry breaking pattern of the system was considered by taking $\kappa_{1}>\kappa_{1}^{\textrm{crit}}$ to be fixed and increasing $\kappa_{2}$ across its critical value, \textit{i.e.}, the transition (C) to (D), corresponding to the global symmetry breaking pattern $O(3)\times \mathbb{Z}_{2} \rightarrow O(2)$.
This is shown in \cref{fig:bet = 5.5; k1=0.133; et1=0.003; et2=0.001; mu=0.0; xi1=0.0001; xi2=0.00005; xi3=0.00005;xi4=0.0; xi5=0.0}.
Since now it is the VEV from $\Phi_{1}$ responsible for the $W$-boson mass in the (C) phase, the mass from $W_{11}^{j}$ corresponds to the gauge boson mass.
$W_{22}^{j}$ couples to a scattering state for $\kappa_{2}<\kappa_{2}^{\textrm{crit}}$, and to the gauge vector after the second scalar passes to its broken phase.
This is clear in the left plot, with the masses from $W_{11}^{j}$ and $W_{22}^{j}$ matching in the (D) region.

For the scalar sector in the region (C) the operator $S_{11}$ now represents the SM Higgs, $m_{h}$, while the spectrum of $S_{22}$ starts at $2m_{H}$.
When crossing to (D) the continuous group $O(3)$ is broken down to $O(2)$ generating two pseudo-Goldstone bosons.
These couple to the mixed operator $W_{12}^{4}$ \cite{maas_gauge_2016}, which can be seen to have a very small mass in the (D) region.

\begin{figure}[htb!]
  \centering
      \begin{minipage}[t]{.5\textwidth}
		\centering
		\scalebox{0.5}{\includegraphics[scale=0.7]{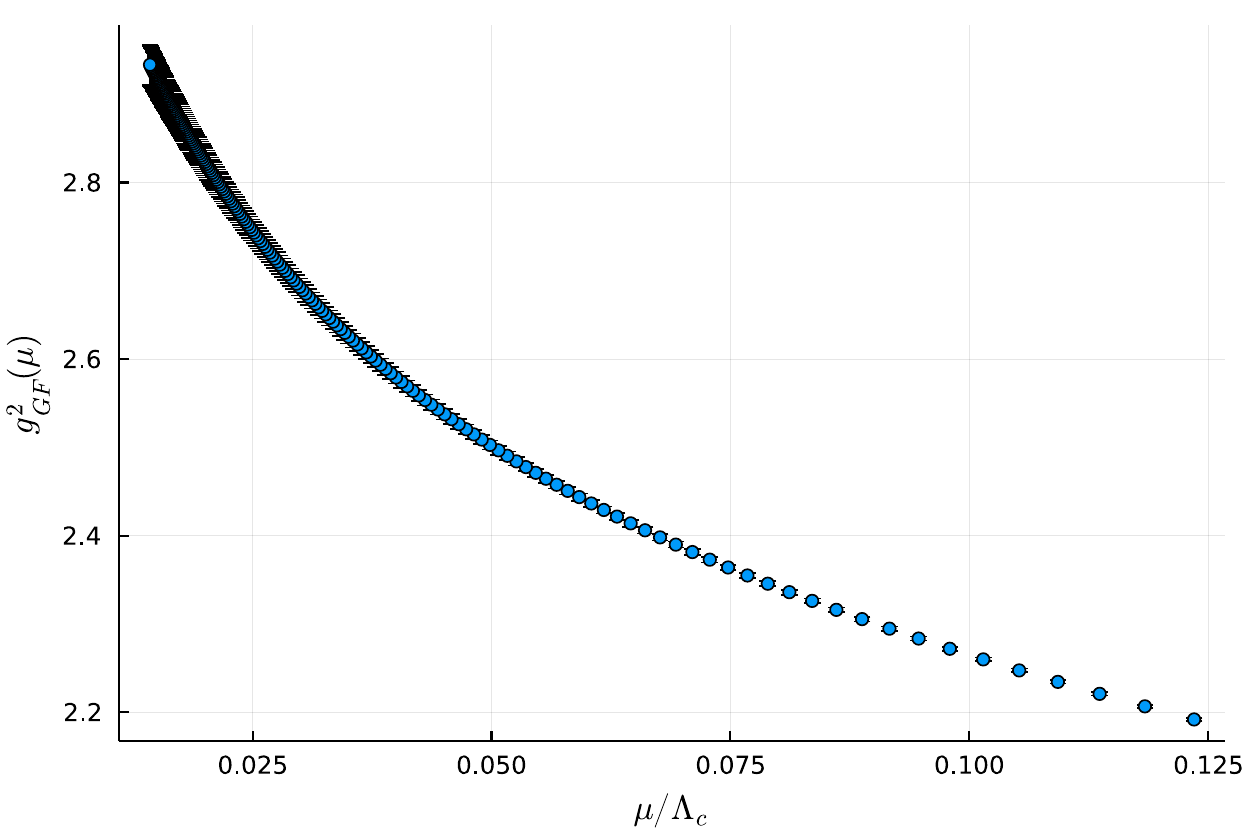}}
	\end{minipage}%
    \begin{minipage}[t]{.5\textwidth}
		\centering
		\scalebox{0.5}{\includegraphics[scale=0.71]{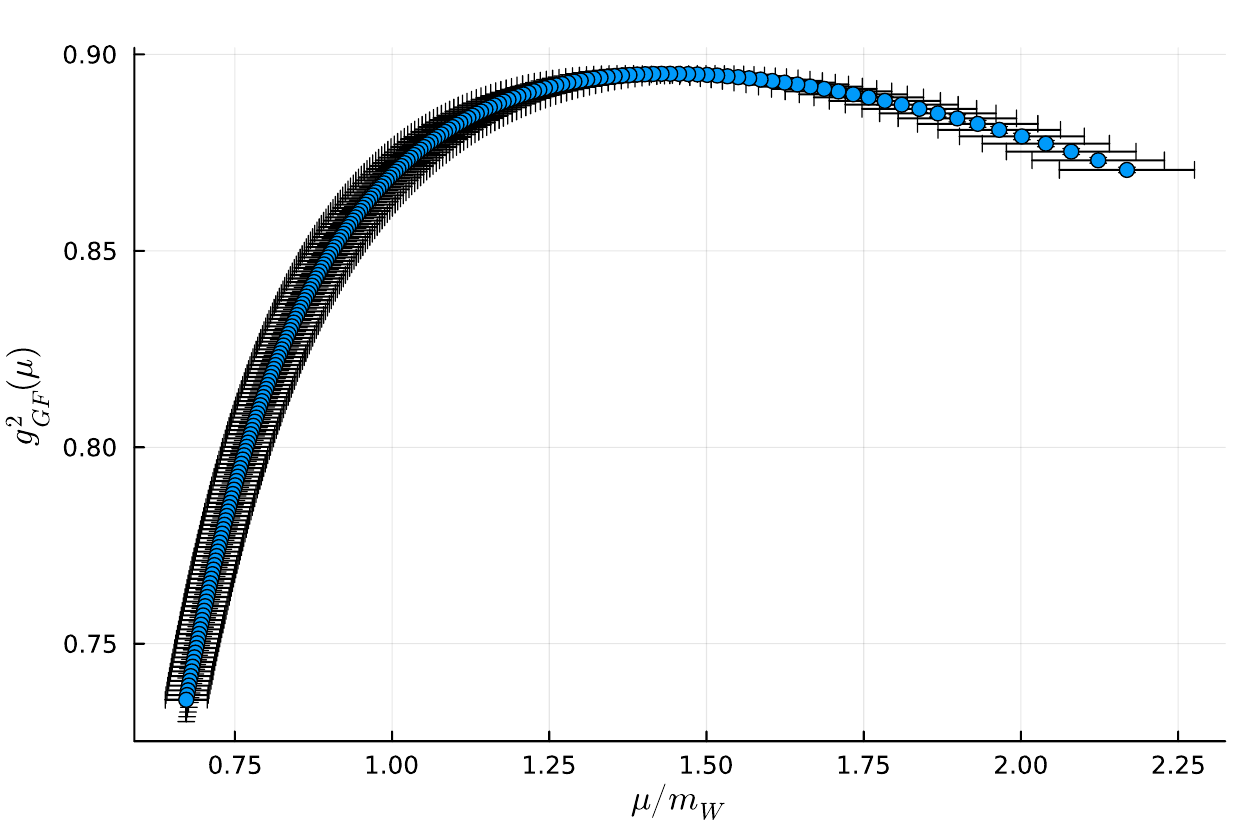}}
	\end{minipage}%
    \caption{Gradient flow running gauge coupling in the confining phase, $\beta=3.6, \kappa_{2}=0.1, \eta_{2}=0.001$ (left) and in the Higgs phase $\beta=6.0, \kappa_{2}=0.1322, \eta_{2}=0.004$ (right). The former is plotted as a function of the dimensionless ratio $\mu/\Lambda_{c}$, $\Lambda_{c}=1/a$ is the lattice cutoff, due to the absence of the the W mass scale. In the Higgs phase the massive W provides a reference scale and the running coupling is shown as a function of $\mu/m_{W}$.}
	\label{fig:running coupling}
\end{figure}

While at this stage of the work we have not fixed the running gauge coupling to its physical value, we already performed an extensive investigation of its behavior in different phases.
In \cref{fig:running coupling} the gradient flow running gauge coupling, $g_{GF}^{2}(\mu)$, is shown as a function of the energy scale in lattice units $\mu=1/\sqrt{8t}$ for the symmetric phase (left-plot) and as a function of the dimensionless scale $\mu/m_{W}$ in the broken phase (right-plot).

In the confining phase the running coupling shows the usual QCD-like behavior with the coupling growing for lower energies.
On the other hand, the scale dependence in the Higgs phase follows a different pattern.
For energies high above the $m_{W}$ scale the gauge boson is effectively massless.
Consequently, the gauge running coupling increases as we lower the scale.
When the energy scale approaches $m_{W}$ from above, the mass of the W becomes relevant and the coupling stops increasing.
In the opposite end, for energies far below $m_{W}$, the gauge boson decouples and we effectively recover a scalar theory.
This shows qualitative agreement with perturbative predictions, which gives the one-loop $\beta$-function of this theory as
\begin{equation}
    \beta_{SU(N)+\rm Scalars}= \mu \dv{g}{\mu} = -\frac{b_0g^3}{16\pi^2}+\order{g^5},~~b_0=\frac{11N-n_s}{3}.
\end{equation}
were $n_{s}$ is the number of scalar fields.
Since the gauge bosons are integrated out when $\mu$ is below $m_{W}$, effectively $N$ can be set to zero and the above one-loop $\beta$-function becomes positive in this regime.

\section{Conclusion} \label{sec:Conclusion}

The 2HDM's are described by a large set of couplings allowing for various scenarios categorised by global symmetry breaking patterns.
All of these require a thorough non-perturbative study in order to identify the ones that are physically realizable.
These global symmetry breaking patterns affect the spectrum of the theory, resulting in different mass hierarchies amongst the low-lying scalar and vector states.
We have set up the machinery that allows us to explore the most general 2HDM scalar potential with real couplings through lattice simulations.

As a first test we consider the custodial symmetry limit and the vanishing of the $\mathbb Z_{2}$-breaking terms $\mu^{2},~\eta_{6}$, and $\eta_{7}$.
We study the behaviour of the phase structure and the spectrum using the perturbative tree-level predictions as a guide.
For the first step of our study, we perform lattice simulations with all the bare couplings being small.
Our results agree qualitatively with predictions from perturbation theory, as expected.
The identification of the physical spectrum with the interpolators for small quartic couplings requires a careful analysis of the tree-level mass-matrices for different global symmetries of the scalar potential.
In this work we consider only cases where the matrices are diagonal, but already at the perturbative level the most general scalar potential generates mixing between the scalar fields.
Additionally, since in our future work we aim at exploring the effect of having $\order{1}$ quartic couplings in the $\mathbb Z_{2}$ broken theory ($\mu^{2},~\eta_{6}$, and $\eta_{7}$ not vanishing), a variational approach is required.
We must solve the generalised eigenvalue problem in order to disentangle the contributions of the mass eigenstates from the mixture of scalar states.

Lastly, the running gauge coupling is computed for both phases using the gradient flow action density, and its  behaviour agrees qualitatively with that predicted by one-loop perturbation theory.

The goal of this exploratory study is to demonstrate that we are able to tune the bare couplings to reproduce the physical Higgs-to-W mass ratio, as well as the value of renormalised gauge coupling at the scale $m_{W}$.
In the future we aim to obtain the spectrum of the theory on finer lattices while following the line of constant physics.
While a proper continuum extrapolation cannot be performed due to triviality, this will allow us to understand the spectrum of the theory and its relation to the cut-off scale, as well as how the diverse scenarios may adjust to the experimental results.

\section*{Acknowledgements}

The authors acknowledge financial support from the Generalitat Valenciana (genT program CIDEGENT/2019/040), Ministerio de Ciencia e Innovacion (PID2020-113644GB-I00), the Academic Summit Project -- NSTC 112-2639-M-002-006-ASP of Taiwan.
GC, CJDL, and AR acknowledge the CSIC-NSTC exchange grant 112-2927-I-A49-508.
CJDL acknowledges the financial support from NSTC project 112-2112-M-A49-021-MY3.
Similarly, GWSH, CJDL and MS acknowledge the support from the NSTC project 112-2639-M-002-006-ASP.

\bibliography{references}

\end{document}